\documentclass[10pt,conference]{IEEEtran}
\IEEEoverridecommandlockouts
\PassOptionsToPackage{table,xcdraw}{xcolor}
\usepackage{tcolorbox}
\usepackage{fontawesome}
\usepackage{booktabs}
\usepackage{cite}
\usepackage{amsmath,amssymb,amsfonts}
\usepackage[capitalise]{cleveref}
\usepackage{algorithmic}
\usepackage{graphicx}
\usepackage{textcomp}
\usepackage{float}
\usepackage{xurl}
\usepackage{flushend}
\usepackage{tikz}
\usepackage{pgfplots}
\pgfplotsset{width=10cm,compat=1.9}

\def\BibTeX{{\rm B\kern-.05em{\sc i\kern-.025em b}\kern-.08em
    T\kern-.1667em\lower.7ex\hbox{E}\kern-.125emX}}
\begin{document}
\definecolor{darkgreen}{rgb}{0, 0.7, 0}
\definecolor{darkred}{rgb}{0.7, 0, 0}
\title{Time Warp: The Gap Between Developers’ Ideal vs Actual Workweeks
in an AI-Driven Era}

\author{
\IEEEauthorblockN{Sukrit Kumar\IEEEauthorrefmark{1}\IEEEauthorrefmark{2}, Drishti Goel\IEEEauthorrefmark{4}\IEEEauthorrefmark{2}, Thomas Zimmermann\IEEEauthorrefmark{7}, Brian Houck\IEEEauthorrefmark{4}\IEEEauthorrefmark{6}, B. Ashok\IEEEauthorrefmark{5}\IEEEauthorrefmark{6}, Chetan Bansal\IEEEauthorrefmark{4}\IEEEauthorrefmark{6}}
\IEEEauthorblockA{\IEEEauthorrefmark{1}\textit{Georgia Institute of Technology}, \IEEEauthorrefmark{4}\textit{Microsoft}, \IEEEauthorrefmark{5}\textit{Microsoft Research},
\IEEEauthorrefmark{7}\textit{University of California, Irvine}}
%\IEEEauthorblockA{sukritk@gatech.edu, t-drgoel@microsoft.com, tzimmer@microsoft.com, \\ brian.houck@microsoft.com, bash@microsoft.com, chetanb@microsoft.com}
\thanks{\IEEEauthorrefmark{6} Contacts: chetanb@microsoft.com, brian.houck@microsoft.com,\newline bash@microsoft.com.}
\thanks{\IEEEauthorrefmark{2} Sukrit Kumar and Drishti Goel are joint first authors of this work.}

% \IEEEauthorblockN{Sukrit Kumar\thanks{Joint first author}}
% \IEEEauthorblockA{\textit{Georgia Institute of Technology} \\
% sukritk@gatech.edu}
% \and
% \IEEEauthorblockN{Drishti Goel\thanks{Joint first author}}
% \IEEEauthorblockA{\textit{Microsoft} \\
% t-drgoel@microsoft.com}
% \and
% \IEEEauthorblockN{Thomas Zimmermann}
% \IEEEauthorblockA{\textit{Microsoft Research} \\
% tzimmer@microsoft.com}
% \and
% \IEEEauthorblockN{Brian Houck}
% \IEEEauthorblockA{\textit{Microsoft} \\
% brian.houck@microsoft.com}
% \and
% \IEEEauthorblockN{B. Ashok}
% \IEEEauthorblockA{\textit{Microsoft Research} \\
% bash@microsoft.com}
% \and
% \IEEEauthorblockN{Chetan Bansal} 
% \IEEEauthorblockA{\textit{Microsoft} \\
% chetanb@microsoft.com}
}

\maketitle

\begin{abstract} 
Software developers balance a variety of different tasks in a workweek, yet the allocation of time often differs from what they consider ideal. Identifying and addressing these deviations is crucial for organizations aiming to enhance the productivity and well-being of the developers.
In this paper, we present the findings from a survey of 484 software developers at Microsoft, which aims to identify the key differences between how developers would like to allocate their time during an ideal workweek versus their actual workweek. Our analysis reveals significant deviations between a developer's ideal workweek and their actual workweek, with a clear correlation: as the gap between these two workweeks widens, we observe a decline in both productivity and satisfaction. By examining these deviations in specific activities, we assess their direct impact on the developers' satisfaction and productivity. Additionally, given the growing adoption of AI tools in software engineering, both in the industry and academia, we identify specific tasks and areas that could be strong candidates for automation. 
In this paper, we make three key contributions: 1) We quantify the impact of workweek deviations on developer productivity and satisfaction 2) We identify individual tasks that disproportionately affect satisfaction and productivity 3) We provide actual data-driven insights to guide future AI automation efforts in software engineering, aligning them with the developers' requirements and ideal workflows for maximizing their productivity and satisfaction.    
% We provide actionable insights for teams and organizations to optimize their workflows to boost developer productivity and also for researchers and developers working on building AI-driven tools for software engineering, helping focus on areas that might lead to greater impact.
% Our goal is to contribute to the ongoing AI-automation efforts by aligning them with developers' perspectives on their ideal workflows, ultimately providing insights to enhance both productivity and satisfaction. 

\end{abstract}


\section{Introduction}

In software engineering, the productivity and satisfaction of developers are pivotal factors that influence both individual performance, customer experience and ultimately, organizational success ~\cite{GRAZIOTIN201832,storey2019towards}. The day-to-day activities which define a developer's workweek encompass a broad spectrum of tasks; from coding and designing new systems, to preparing documents, attending meetings, on-boarding new employees, adhering to security and compliance tasks, etc \cite{8666786}. Each of these tasks is integral to the software development life cycle. Ideally, developers would prefer to allocate their time across these tasks in a way that optimizes both productivity and satisfaction— this can be referred to as their `ideal workweek'. However, in practice, their `actual workweek', can vary significantly from their `ideal' due to fluctuating workloads, shifting organizational priorities, dependencies on other teams, technical challenges, the influence of the work environment, etc \cite{10.5555/319568.319651}, \cite{10.5555/782010.782031}, \cite{300082}.

This misalignment between the ideal and actual workweek raises several intriguing research questions. Key among them is how these deviations impact developers’ productivity and satisfaction. By examining these deviations, we can uncover patterns that may not only reveal  the activities where developers struggle the most but also where they feel most productive and satisfied. Understanding these dynamics provides actionable insights for engineers, managers and organizations to create more effective work environments.

Furthermore, gaining deeper insights into the relationship between time allocation across activities, productivity, and satisfaction could inform the development of future and current AI-driven automation tools. 
With the growing role of AI in software engineering, understanding which tasks developers are most eager to automate becomes crucial.
For instance, certain tasks may seem to offer high productivity but are inherently dissatisfying for developers to engage with. Identifying these tasks can help guide future research and development of these tools to be better aligned with real-world needs of the developers, allowing them to focus on more satisfying work. 

% \sukrit{Change a few lines here. See latex for original} 
% tasks that may give a sense of high productivity, but are actually dissatisfying to work on— could help guide research and development efforts to align with real-world needs, allowing developers to focus on more creative and intellectually engaging work. 

In this paper, we present the findings from our study aimed at analyzing the differences between a developer's `ideal' and `actual' workweek.
Our study builds upon existing research in this field, particularly those exploring developers' typical and optimal workdays \cite{8666786}. By analyzing deviations at the weekly level, we capture the cumulative effects of task allocation, while providing a more fine-grained understanding of how various activities impact productivity and satisfaction.
To the best of our knowledge, this study is the first work that focuses on how deviations, both at the overall workweek and individual activity levels, affect developers' productivity and satisfaction. Our research additionally quantifies how deviations across different activities impact self-reported developer productivity and satisfaction. We also explore how AI tool usage influences the productivity and satisfaction of developers. In addition to that, we systematically analyze and present the developers' feedback on which tasks they would prefer to see automated using AI, providing valuable insights for future AI tool development. 

An outline of the key research questions driving this study are as follows:
\begin{itemize}
    \item[\textbf{RQ1:}] How do developers allocate their time during a typical workweek, and how does this compare to their perception of an \textbf{ideal workweek?}
    \item[\textbf{RQ2:}] How are developer's satisfaction and productivity affected by \textbf{deviations} from their ideal workweek?
     \item[\textbf{RQ3:}] For which tasks do developers prefer using \textbf{AI tools}, and how does the frequency of AI tool usage \textbf{influence} their satisfaction and productivity?
\end{itemize}

The remainder of the paper is structured as follows. In Section \ref{sec:methodology}, we describe our survey design and limitations. In Section \ref{sec:comp_ideal} we perform an analysis on the developers' ideal and actual workweeks. In Section \ref{sec:deviation_impact}, we analyze the impact of deviation from an ideal workweek on developer productivity and satisfaction. Section \ref{sec:ai_tools}  discusses the impact of AI tool usage on developers and identifies activities that developers want to automate. In Section \ref{sec:discussion}, we discuss the implications of our findings for developers, managers and organizations. In Section \ref{sec:related_work}, we discuss related works and, we finally present the conclusion in Section \ref{sec:conclusion}.

\section{Methodology}
\label{sec:methodology}

In this section, we provide a detailed description of the methodology used to design and conduct the survey. We then discuss potential limitations of our study.

\subsection{Survey Design}

To gain insights into the types of activities developers engage in during a typical workweek, we first conducted a series of exploratory interviews with 12 randomly selected participants. These semi-structured exploratory interviews provided a qualitative foundation, allowing us to iteratively develop a comprehensive list of high-level activities that reflect what a developer does over the course of a workweek. The findings from these interviews were instrumental in refining our survey design.

The survey was distributed to software engineers working in Microsoft teams across India and the United States between June to July 2024. A total of 6000 randomly selected individual contributor (IC) developers were invited to participate in the survey via email across multiple batches. We also sent follow-ups to people who had started the survey but had not finished it and also to those who hadn't started it. We framed it as a study aimed at boosting developer productivity by helping us understand how they allocate their time in a workweek. The survey received 484 complete responses (a response rate of 8.06\%). In the invite, the participants were also informed that they could enter a sweepstake to win one out of ten \$50 Amazon.com Gift Cards after finishing the survey.
The ethics for this survey were reviewed and approved by the Microsoft Research Institutional Review Board (MSR-IRB), which is an IRB federally registered with the United States Department of Health \& Human Services. 

\begin{table}[ht]
    \centering
    \caption{Key Activities identified during the exploratory interviews}
    \renewcommand{\arraystretch}{1.2} 
    \begin{tabular}{>{\centering\arraybackslash}p{0.8\linewidth}} 
    \toprule
    \rowcolor[HTML]{E7E7E7} \textbf{Developer Activities} \\
    \midrule
    Architecting and designing new systems \\
    Coding new features \\
    Development Environment Setup \\
    Documentation \\
    Pull Requests/Code Review \\
    Code Refactoring \\
    Debugging during development \\
    Setting-up Monitoring and Dashboards \\
    Authoring Tests \\
    Security and Compliance \\
    Addressing Customer Support Tickets \\
    Communication and Meetings \\
    Task Creation and Management \\
    Giving Technical Presentations \\
    Mentoring and Onboarding \\
    Learning New Technologies \\
    \bottomrule
    \end{tabular}
    
    \vspace{0.5cm}
    \label{tab:developer_activities}
\end{table}

% Find a better way to include these activities

From the exploratory interviews, we identified sixteen key high-level activities, which were subsequently used to quantify the developers' time allocation in the survey. The list of activities are listed out in Table \ref{tab:developer_activities}.
We designed the survey questions to broadly cover the main research questions that we had in mind and also included additional questions which we thought could offer interesting new insights. A brief summary of the questions is given below:
\begin{itemize}
    \item {Their current role and years of experience in the industry and team}
    \item {The number of hours spent on various activities in the previous workweek}
    \item{ The percentage of time they would want    to allocate to each activity in an \emph{ideal} workweek}
    \item {How productive and satisfied they felt in the past workweek}
    \item {Activities they found to be most cognitively challenging} 
    \item {How often do they use AI tools to assist them in various activities}
    \item {Two open-ended questions about the tasks they would want to automate using AI tools, and advice for new hires to boost their productivity and satisfaction levels} 
\end{itemize}

% \sukrit{Remove if we don't add this}
% The list of activities with their description and questions are available here ~\cite{}.

\subsection{Limitations}

This study presents an exploratory survey, which, to the best of our knowledge, is the first of its kind to introduce the concept of a developer's perceived ideal workweek and assess how deviations from it impact productivity and satisfaction. Additionally, it explores the specific activities developers would prefer to automate using AI tools. While this research provides valuable insights, the following limitations should be addressed by future studies. 

\textbf{External Validity:} Since the study was conducted within a single company, the results may not be generalizable to organizations of different sizes, structures, or roles. Additionally, the low response rate and the survey's confinement to specific teams further limit the generalizability of the findings to developers in other contexts.

\textbf{Construct Validity:} The survey relies on the developers to accurately self-report the time they spend on the key activities, productivity and satisfaction levels, which may introduce inaccuracies and biases. Additionally, the sequence of questions could have influenced the responses (e.g. activities to automate)

\textbf{Internal Validity:} The key activities identified in the survey may vary across different developers and organizations. While we sought to minimize this variability through exploratory interviews, the findings may still differ for other developer populations. Additionally, the survey data was collected during a security push at Microsoft, which may have impacted the typical workweek. To mitigate this, we distributed the survey in batches over a month and asked developers to report time spent on activities over the last five workdays, helping to average out short-term fluctuations from the security push and feature releases. 

% One possible limitation of the study is that the results might not translate to other companies of different sizes and shapes. Given that the survey was conducted at Microsoft. In addition to this another limitation was the inability to verify the time that the participants entered to the actual time. 

% \sukrit{Re-write the following lines: } The data for the survey was collected during an ongoing security push inside of Microsoft. So, there is a possibility that the typical workweek during this period might be more different than a workweek during a regular time. However, we tried to balance for this by sending our survey in batches over a month allowing us to reduce the effect of this. We also explicitly asked developers for the time spent on activities in the last 5 working days. This would allow us to average out short term variations due to the security push and feature releases etc.

\section{Comparing Developer's Actual vs. Ideal Workweek}
\label{sec:comp_ideal}

To answer RQ1, we asked the developers to specify the number of hours that they spend on key activities during a typical workweek, as well as the percentage of time they would ideally allocate to these activities. We transform the actual workweek responses into the percentage scale by dividing the time spent on each activity by the total number of hours in the response. We present the developers' actual and ideal workweek responses in \cref{fig:workweekBOX}(a) and \cref{fig:workweekBOX}(b) respectively, and also highlight the key differences in the percentage of time spent between them in \cref{fig:A_vs_I}.

\subsection{The Actual Week: How Developers Spend Their Time}

In the actual workweek, as depicted in \cref{fig:workweekBOX}(a), there is a considerable variation in the time developers spend across activities, specifically for `Coding', `Architecting \& designing new systems', `Security \& Compliance', `Debugging', and `Communication \& Meetings', with wide inter-quartile ranges and the presence of many outliers. This could be attributed to the differences in the type of project, team structure, level of experience, or individual roles. 
Furthermore, developers dedicate the majority of their time on `Communication \& Meetings' ($\approx$12\%), `Coding' ($\approx$11\%), `Debugging' ($\approx$9\%), `Architecting and designing new systems' ($\approx$6\%), and `Pull Requests/Code Reviews' ($\approx$5\%). Conversely, they spend the least time on `Learning New Technologies', `Mentoring and Onboarding', `Giving Technical Presentations', and `Setting-up Monitoring and Dashboards'. 

% Interestingly, the fact that `Communication \& Meetings' consume the largest share of developers' time in the actual workweek. 
% may help explain the findings from the study conducted by Houck \textit{et al.}\cite{houck2023the}, which identified ``Too many meetings" as the second most frequently cited workplace challenge.

\subsection{The Ideal Week: How Developers Want to Spend Their Time}
\begin{figure}
    \centering
    \includegraphics[width=\linewidth]{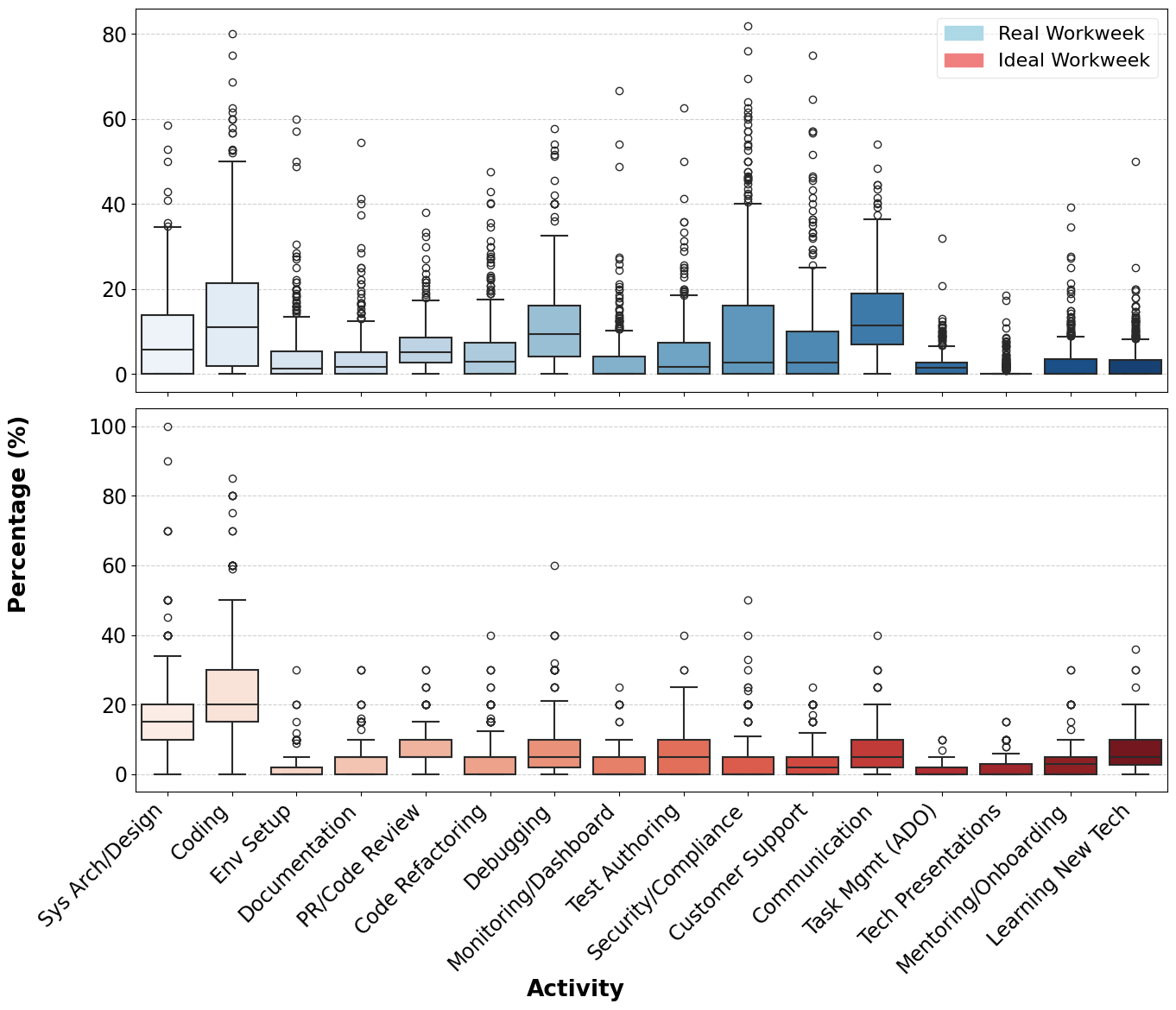}
    \caption{Box Plot of Percentage Time Spent in an Actual Workweek and Ideal Workweek. (a) Top plot: Real Workweek. (b) Bottom plot: Ideal Workweek.}
    \label{fig:workweekBOX}
\end{figure}

% \sukrit{Can we compute the variance per distribution ? (ignore if done later)} 
In contrast to the actual workweek, the ideal workweek shows a less varied distribution across most activities, as seen in \cref{fig:workweekBOX}(b). Developers prefer to allocate their time primarily to `coding' ($\approx$ 20\%) and `Architecting \& designing new systems' ($\approx$ 15\%), while maintaining a more balanced distribution across the remaining activities.
This desire for balance is echoed by individual developers in their personal experiences. For instance, developer D223 offers this advice to new employees: \textit{``Maintain a good balance of all the activities needed to prevent burn out, and start with the activities that you find interesting."}

% \begin{figure}[H]
%     \centering
%     \includegraphics[width=\linewidth]{Sections/Images/SEIP_IdealWW.png}
%     \caption{Box Plot of Percentage Time Spent in an Ideal WorkWeek}
%     \label{fig:IdealWW}
% \end{figure}

The most striking differences are observed in `Communication \& Meetings' and `Task Creation \& Management', where developers show a strong preference for reducing the time spent. These activities are marked by high variability in the actual workweek, but are more contained in the ideal scenario, reflecting a widespread desire to minimize the time spent on these activities. Developer D342 emphasizes this sentiment: \textit{``Optimize for operations. Automate EVERYTHING for Continuous Delivery. The word `manual' should not be in your vocabulary unless referring to something to read. This way, you will have more time to focus on the creative process that is software engineering and be able to release often and with confidence."}. Similarly, D238 adds, \textit{``More focus time and fewer meetings."}

\begin{figure}
    \centering
    \includegraphics[width=\linewidth]{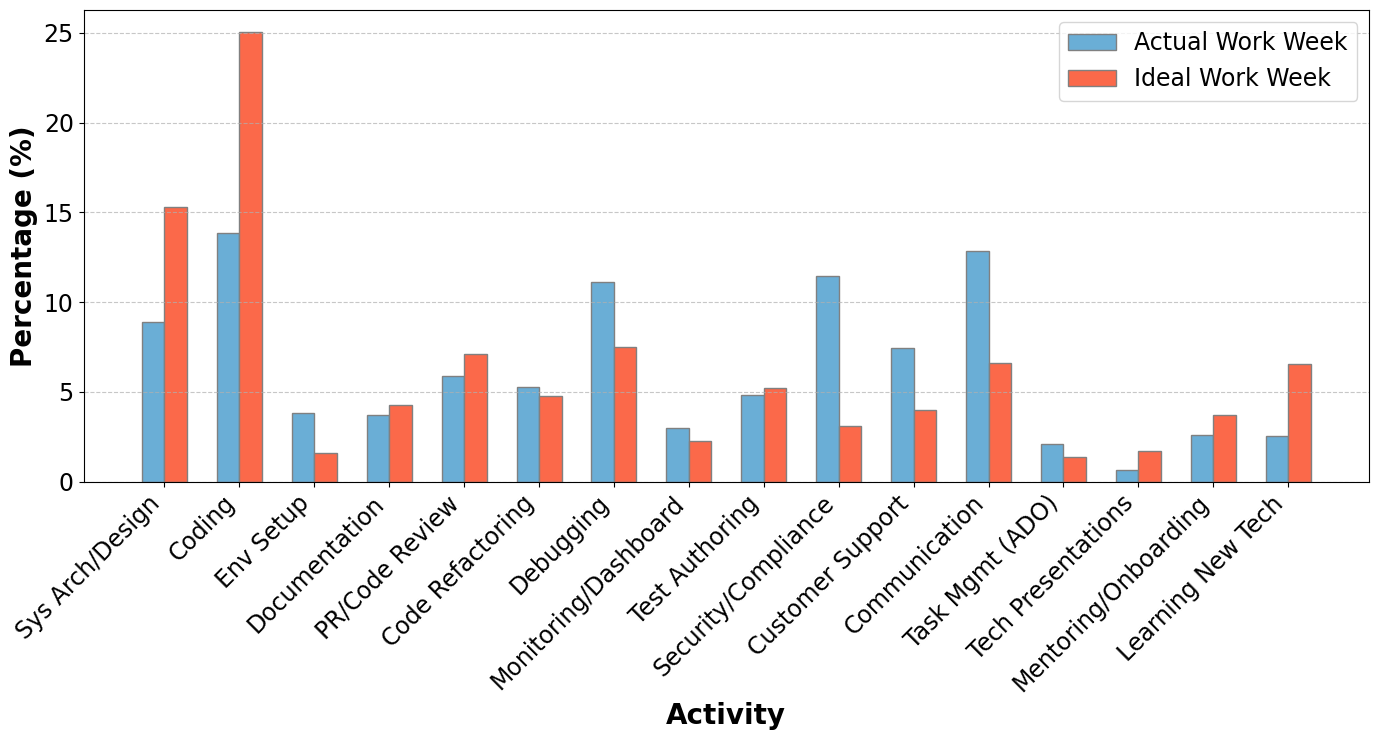}
    \caption{Average percentage of time spent on the key activities in the Actual vs Ideal Workweek}
    \label{fig:A_vs_I}
\end{figure}

In \cref{fig:A_vs_I} we highlight several areas where there is a significant gap between the developers' ideal and actual workweek, shown by the percentage of time spent on each development related activity. 
Developers want to spend more time on core activities such as `Coding', `Architecting and designing new systems', and less time on `Communication', `Addressing Customer Support Tickets', and  `Security \& Compliance'. 
Additionally, the developers desire to spend less time on debugging code, and instead want to invest more time on preventative measures like `PR/Code Reviews', `Documentation', and `Mentoring/Onboarding' that could potentially alleviate pains fixing issues due to poor documentation or code. This indicates a potential desire for improving code quality and reducing long-term maintenance costs and focusing and reducing technical debt~\cite{fowler2018refactoring}. Similarly, the preference to spend more time on learning new technologies, and engaging in technical presentations suggests that developers consider continuous learning and knowledge-sharing an essential part of their workweek. Developer D158 shares their personal experience regarding the importance of continuous learning in a workweek: \textit{``Initially, my learning somewhat slowed as I was too caught up in the urgency to release features. The workweek was about completing as many ADO items as possible. However, my development as an engineer actually came from setting aside blocks of 2-3 hours whenever possible to go deep into certain topics. In my opinion, each team should have a mini boot-camp for onboarding engineers. The material does not need to be fully comprehensive but there should at least be a list of topics that the new member can deeply research. For example, if your job is to connect two internal systems together, you should make it your responsibility to deeply understand both systems even though it's not directly related to your task."}

%Insights box
\begin{tcolorbox}[
    colback=gray!10,  
    rounded corners,  
    boxrule=0.35mm,  
    left=5pt, right=5pt, 
    top=5pt, bottom=5pt 
]

\faLightbulbO\ \textbf{Takeaway}: A developer's perception of an ideal workweek is not well aligned with their actual workweek. Developers want to spend more time on core development activities such as coding, designing new systems, learning new technologies, and knowledge sharing, while wanting to reduce the time spent on dealing with tasks related to security \& compliance, communication, debugging, addressing incident/support tickets, and task management. 
\end{tcolorbox}

\section{Impact of Deviations from the Ideal Workweek on Developer Productivity \& Satisfaction }
\label{sec:deviation_impact}

In this section, we address RQ2 by evaluating how deviations from a developer's self-perceived ideal workweek influence their levels of satisfaction and productivity. Additionally, we identify the specific activity-level deviations and how they contribute to the observed effects.

\subsection{Bridging the Gap: How Deviations from the Ideal Workweek affect Developers' Productivity and Satisfaction}
% Individual-level deviation from the Ideal workweek \& it’s impact on productivity \& satisfaction \ When reality strays
\label{subsec: 4a}

In the survey, we asked the developers to rate their previous week on five different levels of productivity (`Very productive', `Productive', `Neither productive nor unproductive', `Unproductive', and `Very unproductive') and satisfaction (`Very satisfied', `Satisfied', `Neither satisfied nor dissatisfied', `Dissatisfied', and `Very Dissatisfied'). 
We split the developers based on each of these different categories and calculate and plot the Spearman rank correlation coefficient \cite{Spearman2015ThePA}, and mean absolute error \cite{Willmott2005AdvantagesOT} between the actual and ideal workweek, which are defined by the time they spend on the 16 key activities and the time they would ideally allocate to each activity.  

\begin{figure}[t]
    \centering
    \includegraphics[width=\linewidth]{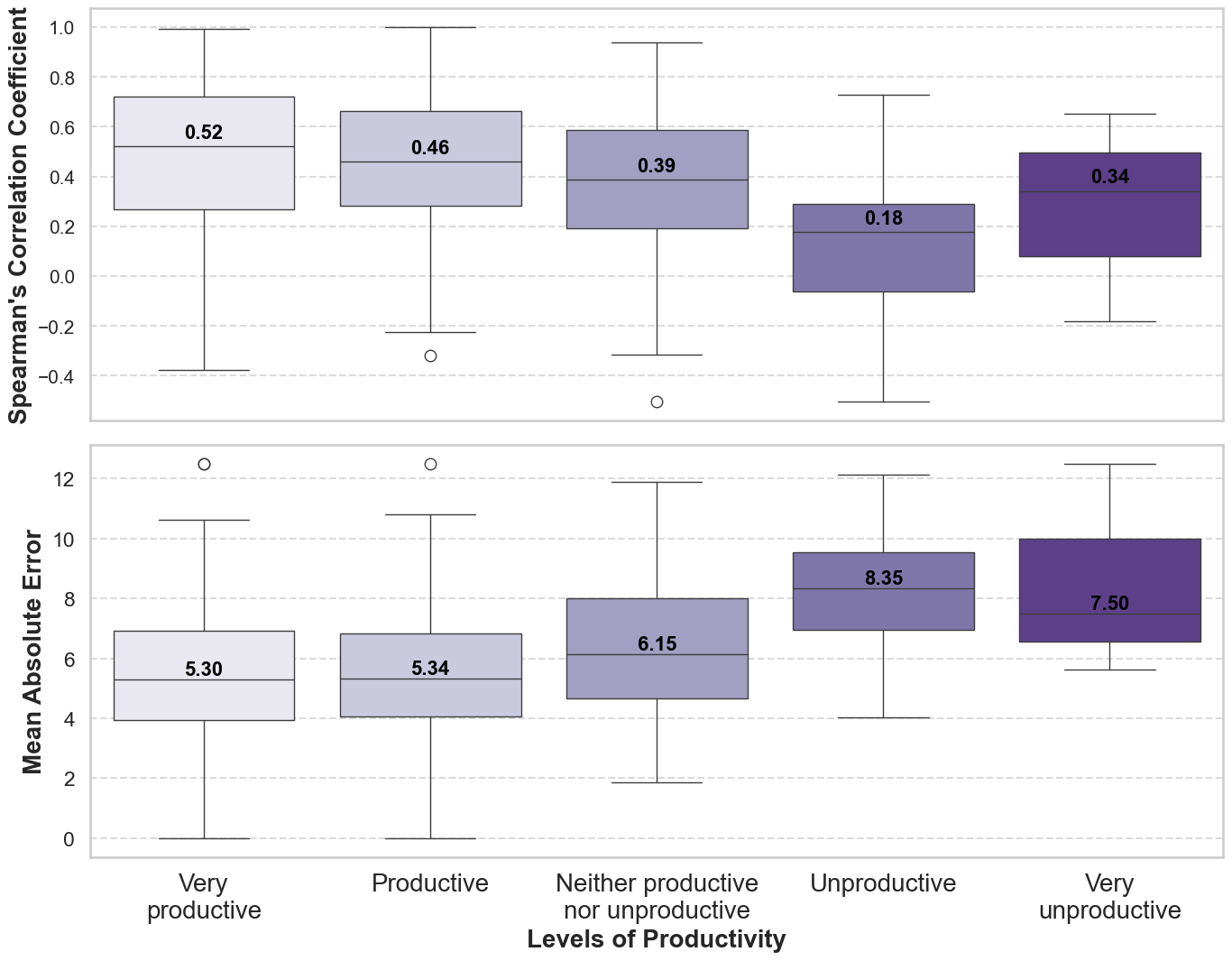}
    \caption{(a) Spearman’s Rank Correlation Coefficient for different categories of productivity and the deviation between the ideal and actual workweek. (b) Mean Absolute Error quantifying the difference in time allocation between the ideal and typical week for different productivity levels.}
    \label{fig: SPM_MAE_Productivity}
\end{figure}

% (a) Spearman's Rank Correlation Coefficient for different categories of Productivity. 
%     (b) Mean Absolute Error for different levels of productivity

\begin{figure}[t]
    \centering
    \includegraphics[width=\linewidth]{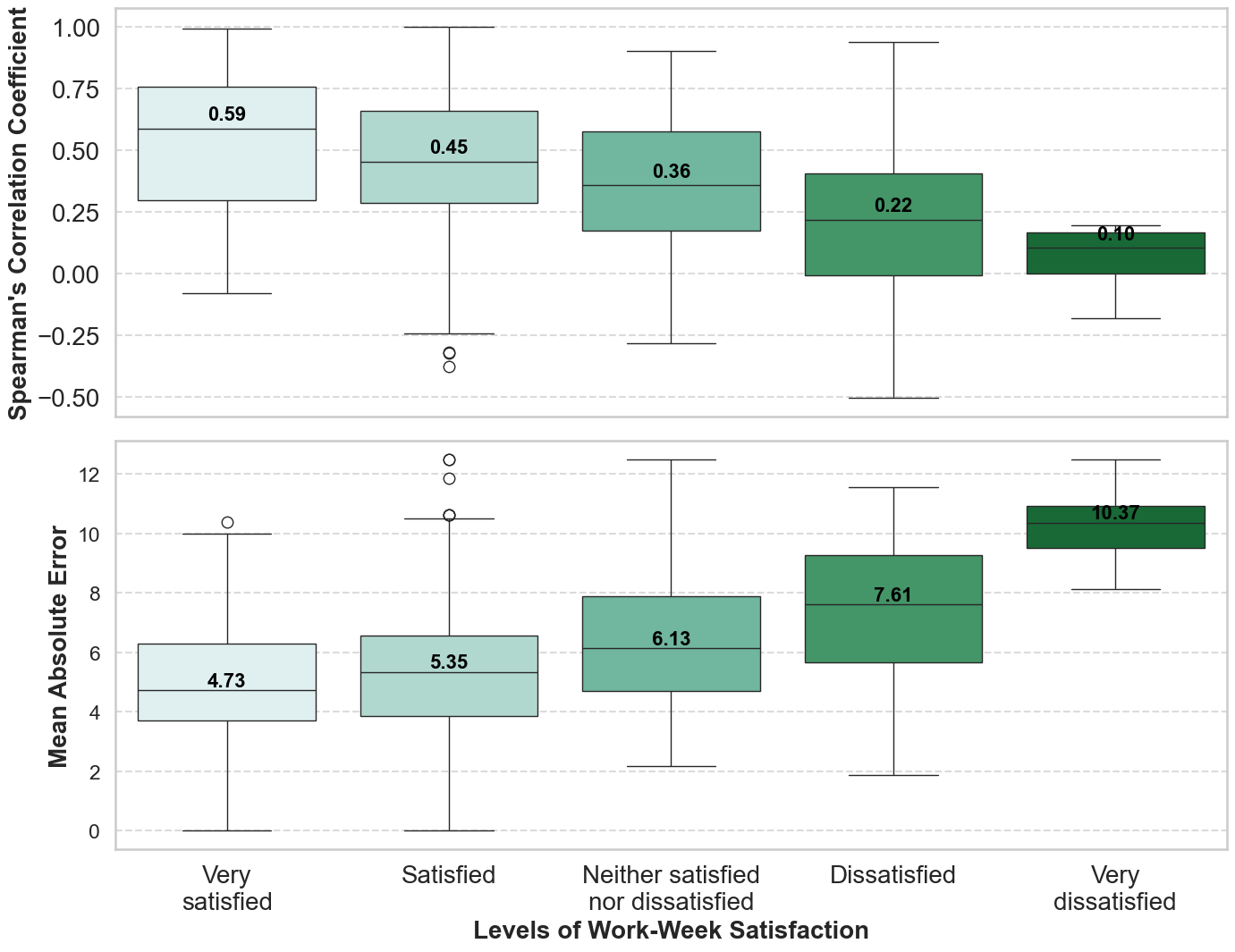}
    \caption{(a) Spearman’s Rank Correlation Coefficient for different categories of satisfaction and the deviation between the ideal and actual workweek. (b) Mean Absolute Error quantifying the difference in time allocation between the ideal and typical week for different satisfaction levels.}
    \label{fig: SPM_MAE_Satisfaction}
\end{figure}

% (a) Spearman's Rank Correlation Coefficient for different categories of Satisfaction. 
%     (b) Mean Absolute Error for different levels of Satisfaction

\cref{fig: SPM_MAE_Productivity}(a) and \cref{fig: SPM_MAE_Productivity}(b), show how differences between actual and ideal workweeks affect developers' productivity. We observe two interesting trends in both the computed metrics. Developers who felt very productive had the highest median Spearman correlation (about 0.52) between their actual and ideal workweeks. This correlation decreased for less productive developers, reaching about 0.18 for unproductive and 0.34 for very unproductive developers. For the mean absolute error (MAE), the difference between actual and ideal workweeks increased as productivity decreased. The MAE rose from 5.3 for very productive developers to 7.5 for very unproductive ones. This shows that the deviation from the ideal work-week is relatively lower for developers who felt productive during a work-week.

% we observe a distinct impact of deviations from the ideal workweek on developers' perceived productivity levels. The median Spearman correlation between the actual and ideal workweek is highest for developers who reported feeling very productive ($\approx$0.52) and declines steadily among those who felt unproductive ($\approx$0.18) and very unproductive ($\approx$0.34). Similarly, the mean absolute error (MAE) between the actual and ideal workweek increases consistently as productivity decreases, rising from 5.3 in the very productive group to 7.5 in the very unproductive group.
% \sukrit{Can we describe the exact method calculation in more detail?} 
% \sukrit{Also can we do a quick bar-plot or text description of the number of devs in each of these 5 buckets ?}
A similar, but more pronounced pattern is observed in the satisfaction levels (\cref{fig: SPM_MAE_Satisfaction}(a) and \cref{fig: SPM_MAE_Satisfaction}(b)). The Spearman's rank correlation between their actual and ideal workweek was again highest for developers who felt very satisfied in that workweek (0.59) and it's lower at 0.10 for those who were very dissatisfied. Likewise, the MAE demonstrates a similar pattern, showing that developers who are more satisfied tend to be closer to their ideal workweek, hence having lower MAE values (4.73 for very satisfied vs. 10.37 for the very dissatisfied group).

%Insights box
\begin{tcolorbox}[
    colback=gray!10,  
    rounded corners,  
    boxrule=0.35mm,  
    left=5pt, right=5pt, 
    top=5pt, bottom=5pt 
]
\faLightbulbO\ \textbf{Takeaway}: As developers' actual workweek diverges from their \textit{perceived ideal workweek}, both productivity and satisfaction show a downward trend. This suggests that aligning with an ideal workweek is not merely a preference, but also crucial for maintaining high levels of performance. It also underscores the necessity of streamlining current research efforts on automation using AI to effectively bridge the gap between the actual and ideal workweek.

%It also underscores the need for research and automation to bridge the gap between the actual and ideal workweek. 

\end{tcolorbox}

\subsection{Impact of activity-level deviations on productivity \& satisfaction}

% \sukrit{Alternate heading ? : Impact of activity-level deviations on productivity \& satisfaction}

In the previous section, we highlighted the impact of deviation from a developer's ideal workweek on their productivity and satisfaction levels.
To gain a more detailed understanding of these deviations, we calculate the differences between the time spent on these activities in the actual and ideal workweek (Actual - Ideal). Each variable in the \cref{tab:regression_results_productivity} and \cref{tab:regression_results_satisfaction} is the difference between the actual and ideal workweek for that particular activity. Positive difference values imply more time being spent on the activity than desired (over-allocation), while negative values imply less time being spent than desired (under-allocation). We apply an Ordinary Least Squares (OLS) regression model to identify specific activities that significantly influence developers' productivity and satisfaction. OLS was chosen for its ability to model linear relationships for continuous variables. Since the primary goal is to assess how incremental changes in activity allocation impact productivity and satisfaction, OLS provides a straightforward, and interpretable framework for the same.
% The general form of the OLS regression equation is as follows:\\
% \begin{equation}
% Y = \beta_0 + \beta_1 X_1 + \beta_2 X_2 + \cdots + \beta_n X_n + \epsilon
% \end{equation}
% where:
% \begin{itemize}
%     \item \( Y \) is the dependent variable (e.g., productivity or satisfaction),
%     \item \( X_1, X_2, \ldots, X_n \) are the independent variables (e.g., differences in time spent on activities),
%     \item \( \beta_1, \beta_2, \ldots, \beta_n \) are the coefficients of the independent variables,
%     \item \( \beta_0 \) is the intercept,
%     \item \( \epsilon \) is the error term.
% \end{itemize}

We highlight activities having statistically significant positive relations in the color green and significant negative relations in red in both the tables.
Conventionally, a p-value of \( < 0.05 \) indicates statistically significant results at a 95\% confidence level, suggesting that the observed deviations in these activities are likely to have a meaningful impact on productivity or satisfaction.

\begin{table}
\centering
\caption{OLS Regression Results - Productivity}
\begin{tabular}{lcccc}
\toprule
\textbf{Variable} & \textbf{coef} & \textbf{std err} & \textbf{t} & \textbf{P}\( > |t| \) \\
\midrule
\textbf{const}                          & 4.1078 & 0.055  & 74.621  & 0.000  \\
\textbf{Arch. \& Design}          & 0.0009 & 0.003  & 0.355   & 0.723  \\
\textcolor{darkgreen}{\textbf{Coding}}                   & \textcolor{darkgreen}{0.0061} & 0.002  & 3.027   & 0.003  \\
\textcolor{darkred}{\textbf{Dev. Environment}}         & \textcolor{darkred}{-0.0158} & 0.005  & -3.422  & 0.001  \\
\textcolor{darkgreen}{\textbf{Docs}}                     & \textcolor{darkgreen}{0.0103} & 0.005  & 1.995   & 0.047  \\
\textbf{PR/Code Review}           & -0.0046 & 0.006  & -0.803  & 0.423  \\
\textcolor{darkgreen}{\textbf{Refactoring}}              & \textcolor{darkgreen}{0.0130} & 0.004  & 2.935   & 0.003  \\
\textbf{Debugging}                & -0.0021 & 0.003  & -0.643  & 0.521  \\
\textbf{Monitoring \& Dashboards} & 0.0063 & 0.005  & 1.197   & 0.232  \\
\textbf{Testing}                  & 0.0045 & 0.004  & 1.093   & 0.275  \\
\textbf{Security \& Compliance}   & -0.0033 & 0.002  & -1.431  & 0.153  \\
\textbf{Customer Support}          & -0.0051 & 0.003  & -1.651  & 0.099  \\
\textcolor{darkred}{\textbf{Communication}}                    & \textcolor{darkred}{-0.0079} & 0.004  & -2.120  & 0.034  \\
\textbf{Task Mgmt.}               & -0.0149 & 0.010  & -1.451  & 0.147  \\
\textbf{Tech-Talks}               & 0.0001 & 0.012  & 0.011   & 0.992  \\
\textbf{Mentoring/Onboarding}     & 0.0007 & 0.006  & 0.119   & 0.905  \\
\textcolor{darkgreen}{\textbf{Learning Tech.}}           & \textcolor{darkgreen}{0.0118} & 0.005  & 2.372   & 0.018  \\
\bottomrule
\label{tab:regression_results_productivity}
\end{tabular}
\end{table}

In \cref{tab:regression_results_productivity} we observe statistically significant positive relations between the time spent on coding, documentation, code refactoring and learning, on developer productivity, while a negative relation for maintaining development environment, and communication. 
Moreover, for satisfaction, \cref{tab:regression_results_satisfaction} highlights a statistically significant positive relation of satisfaction with documentation and learning, while a negative relation for development environment, PR/Code reviews, security and compliance, and communication. 
As described by developer D141 and D18 respectively: \textit{``Keeping organized documentation of project related knowledge will certainly improve productivity"}, \textit{``S360 takes up the majority of my time, which means I spend almost no time creating useful features and responding to customer requests."}
% \sukrit{Do we want to add a takeway for this also ? But would be too many take aways ?}
% \drishti{might be too many...}  I agree lite.

\begin{table}
\centering
\caption{OLS Regression Results - Satisfaction}
\begin{tabular}{lcccc}
    \toprule
        & \textbf{coef} & \textbf{std err} & \textbf{t} & \textbf{P}\( > |t| \) \\
        \midrule
        \textbf{const} & 4.0436 & 0.061 & 66.061 & 0.000 \\
        \textbf{Arch. \& Design} & 0.0057 & 0.003 & 1.950 & 0.052 \\
        \textbf{Coding} & 0.0043 & 0.002 & 1.907 & 0.057 \\
        \textcolor{darkred}{\textbf{Dev. Environment}} & \textcolor{darkred}{-0.0151} & 0.005 & -2.935 & 0.004 \\
        \textcolor{darkgreen}{\textbf{Docs}} & \textcolor{darkgreen}{0.0161} & 0.006 & 2.803 & 0.005 \\
        \textcolor{darkred}{\textbf{PR/Code Review}} & \textcolor{darkred}{-0.0133} & 0.006 & -2.086 & 0.038 \\
        \textcolor{darkgreen}{\textbf{Refactoring}} & \textcolor{darkgreen}{0.0128} & 0.005 & 2.584 & 0.010 \\
        \textbf{Debugging} & -0.0010 & 0.004 & -0.267 & 0.789 \\
        \textbf{Monitoring \& Dashboards} & -0.0098 & 0.006 & -1.692 & 0.091 \\
        \textbf{Testing} & 0.0031 & 0.005 & 0.684 & 0.494 \\
        \textcolor{darkred}{\textbf{Security \& Compliance}} & \textcolor{darkred}{-0.0090} & 0.003 & -3.470 & 0.001 \\
        \textbf{Customer Support} & -0.0065 & 0.003 & -1.900 & 0.058 \\
        \textcolor{darkred}{\textbf{Communication}} & \textcolor{darkred}{-0.0129} & 0.004 & -3.114 & 0.002 \\
        \textbf{Task Mgmt.} & 0.0085 & 0.011 & 0.743 & 0.458 \\
        \textbf{Tech-Talks} & -0.0032 & 0.013 & -0.243 & 0.808 \\
        \textbf{Mentoring/Onboarding} & 0.0077 & 0.007 & 1.142 & 0.254 \\
        \textcolor{darkgreen}{\textbf{Learning Tech.}} & \textcolor{darkgreen}{0.0126} & 0.006 & 2.279 & 0.023 \\
        \bottomrule
    \end{tabular}
    \label{tab:regression_results_satisfaction}
\end{table}

\subsection{Mapping Activity Time Across Productivity-Satisfaction Profiles: A Path Toward Smart Automation}
\begin{figure}[ht]
    \centering
    \includegraphics[width=\linewidth]{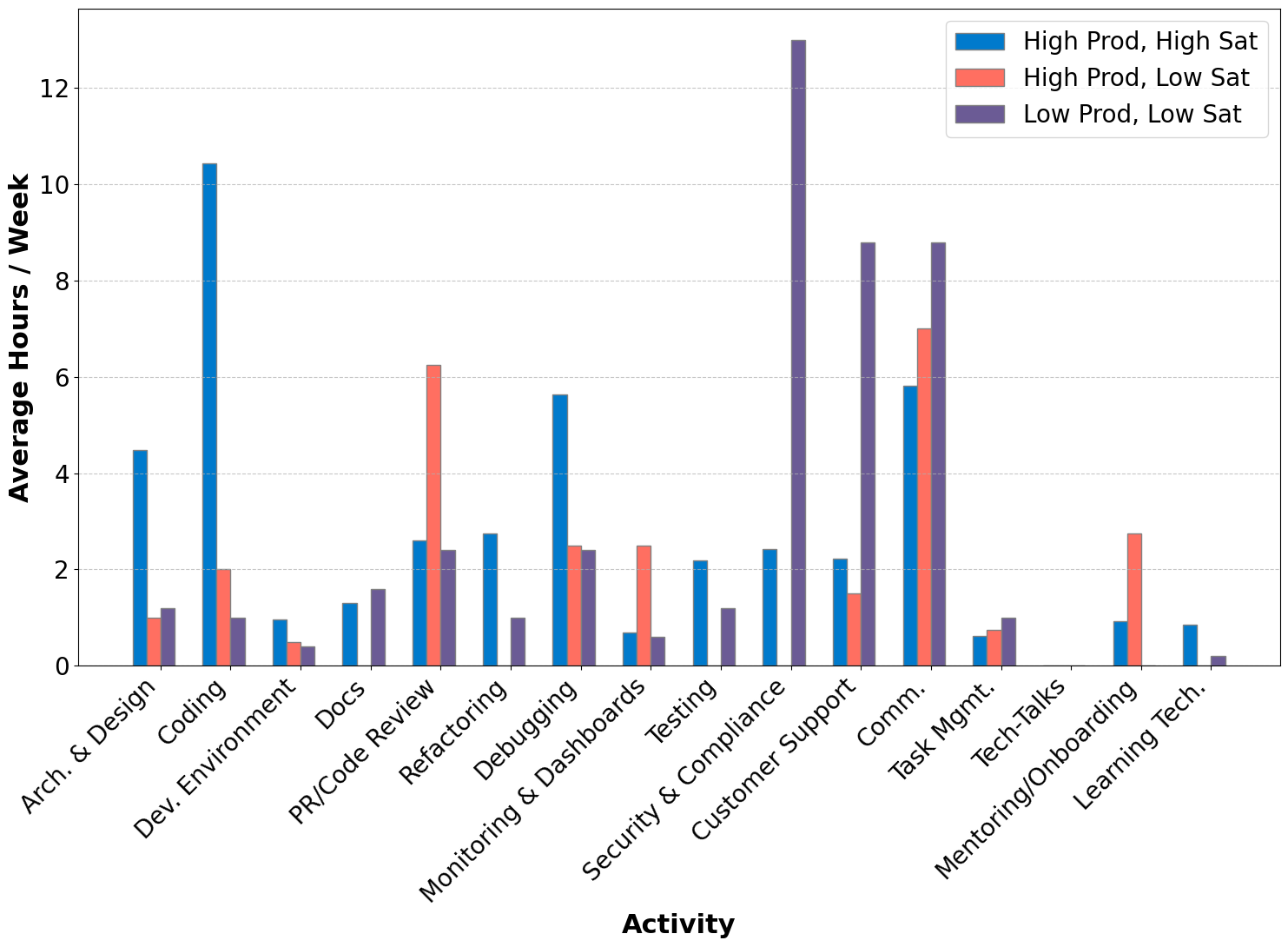}
    \caption{Average actual Weekly Hours Allocated to Various Activities by Developers in High vs. Low Productivity-Satisfaction Groups}
    \label{Hrs_byCategories_ActualWW}
\end{figure}

% Time spent on different activities by participants of different productivity-satisfaction categories (Identifying activities well-suited for automation)
In \cref{Hrs_byCategories_ActualWW}, we categorize developers into three distinct groups: `High Productivity, High Satisfaction,' `High Productivity, Low Satisfaction,' and `Low Productivity, Low Satisfaction,' based on their self-reported productivity and satisfaction levels from the previous workweek (as detailed in \ref{subsec: 4a}). For the sake of simplicity, we merged the `Very Productive' and `Productive' groups into a single `High Productivity' category, and similarly for other categories. The `Low Productivity, High Satisfaction' group was excluded due to its small sample size of two developers. For each remaining group, we visualize the average hours spent on the 16 key activities, highlighting patterns in how time allocation varies across different productivity and satisfaction combinations.

The developers who felt highly productive and satisfied were the ones who spent the most time on activities such as `Architecting \& designing new systems', `Coding', `Debugging', `Refactoring Code', and `Learning New Technologies', while lesser time on `Documentation', `Security \& Compliance', `Addressing Customer Support Tickets', `Communication', `ADO Task Creation and Management' and `Mentoring \& Onboarding'. An opposite trend is observed for the developers from the `Low Productivity, Low Satisfaction' group, as they tend to spend more time on `Security \& Compliance', `Addressing Customer Support Tickets', and `Communication'.

% improved - horiontal

Developers in the `High Productivity, Low Satisfaction' category dedicate most of their time to activities such as `Pull Requests/Code Review,' `Setting up Monitoring and Dashboards,' `Communication,' and `Mentoring and Onboarding.' Despite feeling highly productive during the week, these developers report lower satisfaction levels. This suggests that while they recognize the value of these tasks, they prefer spending less time on them, making these activities ideal candidates for automation. 

%Insights box
\begin{tcolorbox}[
    colback=gray!10,  
    rounded corners,  
    boxrule=0.35mm,  
    left=5pt, right=5pt, 
    top=5pt, bottom=5pt 
]
\faLightbulbO\ \textbf{Takeaway}: 
Developers who experience a sense of productivity, yet feel dissatisfied with their workweek present an excellent opportunity for task automation. Our analysis reveals that this group invests significant time in activities such as `PR/Code Reviews', `Monitoring \& Dashboards', `Communication', and `Mentoring/Onboarding'. Consequently, these tasks emerge as ideal candidates for automation.

\end{tcolorbox}

\section{Impact of AI Tool Usage on Developer Productivity and Satisfaction}
\label{sec:ai_tools}

The rapid advancement of artificial intelligence (AI) and its growing adoption across industries, particularly in automating development tasks, brings forth a crucial question: \textit{Which tasks should be automated using AI to maximize both productivity and satisfaction?} In this section, we observe how the frequency of AI tool usage impacts the developers' productivity and satisfaction and analyze the responses gathered from developers, examining which tasks they believe are most suited for automation.

\subsection{How does the Frequency of AI Tool Usage Influence 
Productivity \& Satisfaction}
In the survey, we asked the developers to indicate how frequently they incorporate AI tools during their workweek, (`Daily', `Weekly', `Monthly', `Once every 6 months', or `Once every year'). We again collapse the top 2 and bottom 2 categories for both productivity and satisfaction into a single category (Productive/Unproductive). 
% The plots in \cref{AI_on_productivity} and \cref{AI_on_satisfaction} show the relationship between AI tool usage frequency and developer productivity and satisfaction. 

\begin{figure}
    \centering
    \includegraphics[width=\linewidth]{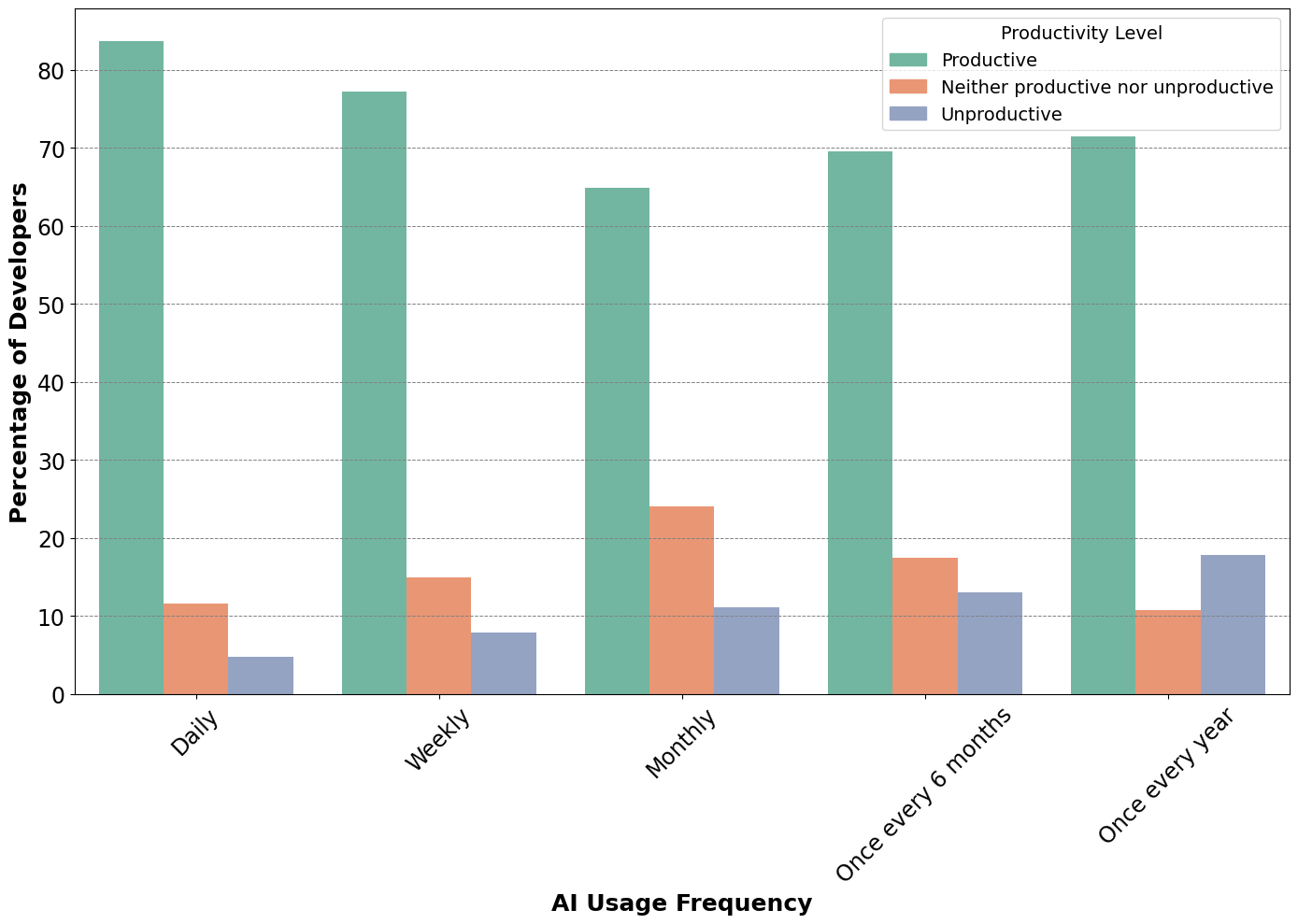}
    \caption{Effect of the frequency of AI tools usage on productivity}
    \label{AI_on_productivity}
\end{figure}

\cref{AI_on_productivity} reveals that developers who use AI tools daily report the highest levels of productivity, with 83.7\% reporting to be ``productive" However, as the frequency of AI tool usage decreases (e.g., from weekly to annually), the proportion of productive developers also declines, while the percentage of those reporting neutrality or even negative productivity outcomes increases.

\begin{figure}
    \centering
    \includegraphics[width=\linewidth]{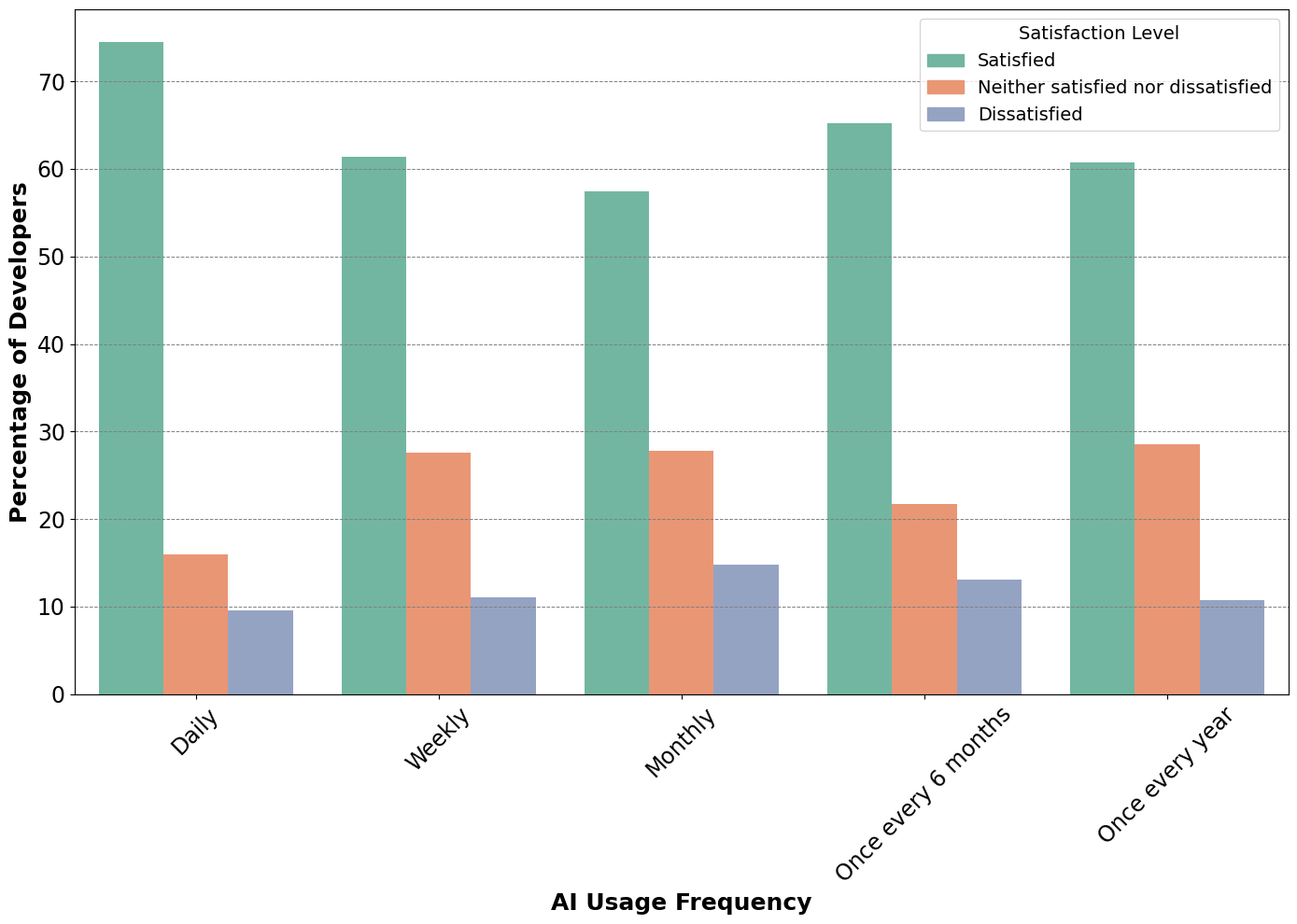}
    \caption{Effect of the frequency of AI tools usage on satisfaction}
    \label{AI_on_satisfaction}
\end{figure}

\cref{AI_on_satisfaction} shows a similar trend between AI tool usage and satisfaction levels. Developers who utilize AI tools on a daily basis experience the highest levels of satisfaction (74.5\%), with most of them reporting to be ``satisfied." Again, as AI usage frequency decreases, dissatisfaction becomes more prevalent.

\subsection{Which tasks do developers want to automate?}

In the survey, we asked the developers to answer the question \textit{``Which of your current tasks would you most like to see automated? Which processes do you think could be enhanced to minimize repetitive work?"} 

We received 242 open-ended responses from the developers. To analyze the responses, we utilized the GPT-4 model to identify an initial set of task categories highlighted in the data. The model identified 17 distinct categories. To ensure comprehensive coverage, we uniformly sampled 60 responses based on their length. Four annotators independently reviewed these responses using a closed-coding approach. 
They assigned labels to each response based on the initial categories suggested by GPT-4, while also identifying any new categories that emerged during the process. 
The inter-rater agreement, measured using the Jaccard similarity metric, was 81.5\%, indicating a substantial level of agreement among annotators. Additionally, through discussions, the annotators agreed to introduce one new category (``Cloud Infrastructure Maintenance''), resulting in a final list of 18 categories of tasks that developers would like to automate using AI. 
The remaining 182 responses were distributed equally amongst the four annotators and manually labelled using the extended taxonomy to identify the frequency of each category.

The final categories are summarized in \cref{tab:AutomationTasks}. For each category, we provide the name and a short description.
The table also illustrates the results from the manual annotation process and the frequency of each category based on the manual annotation results.

Documentation emerged as the most frequent category in our analysis of tasks that could be automated with AI. Developers suggested the use of AI to create and update documentation, maintain team knowledge bases, and also to efficiently understand existing code and systems from the available information. For example, D302 highlighted the importance of efficient documentation by saying: \textit{``I would love to see automated documentation of infrastructure and code. A lot of time is lost trying to understand the current resources that go into securing, building, and deploying our products and trying to understand how the different parts of our code base interact and depend on each other. It'd really lessen the cognitive load if this was well-documented in a way that is automatically kept up-to-date."} Similarly, developer D165 expressed the growing need for automating code documentation to keep pace with the rapid development cycles: \textit{``I would love it if there were some way to automatically generate architectural documentation or either a data flow or call stack diagram based on the structure of the actual code in the repository. While tools like CodeFlow exist for VSCode and Visual Studio, there's a learning curve and the lack of reward for better internal documentation means creating and updating documentation is on a volunteer basis. This, and the rapid pace of development means that onboarding any new engineer or existing engineer to a new area is a time-consuming manual process of last-minute documentation updates, or passing down tribal knowledge verbally. Would be nice if there were some way to tie internal documentation to a particular code base, so that every PR also `refreshes' the internal documentation or diagrams."} 

Following this, several developers emphasized the need for automating tasks associated with the setup and maintenance of development environments, which often involve complex processes such as configuring SSH keys, installing software dependencies, and initializing new development instances. Developer D81 points out the current challenges, stating, \textit{``Setting up the Dev. environment could be way simpler. Right now it is somewhat involved, time consuming, and brittle."} 
Furthermore, D403 emphasizes the significant impact of automation on productivity, noting, \textit{``Setting up a development environment takes up a huge amount of time in the development cycle, hence I feel that enhancing the same will greatly boost developer productivity."}

Developers also suggested the need to automate tasks surrounding the authoring, execution and monitoring of tests. For instance, D81 envisions a future where AI could generate comprehensive acceptance tests based on system usage logs: \textit{``Dreaming wild, what I would like AI to do is to create all the (missing) acceptance tests of the system (maybe by looking at all the usage logs?) because currently any attempt to improve or refactor the code monolith is a risky venture. The only thing that would make it less risky is to have tests that validate that the system functionality is still there."} 

Beyond the aforementioned technologies, developers also convey the importance of enhancing automation efforts to improve various aspects of project management. This includes the efficient creation and management of tasks in systems like Azure DevOps (ADO), which encompasses handling sub-tasks and repeatable tasks, as well as tracking project status effectively. Developer D38 shares his insights on utilizing AI to generate work items in ADO using a structured prompt, such as: \textit{``Create work item with title [Task][GIVE\_TASK\_DESCRIPTION] related to work [Feature\_Item], work description [1 line description] - assigned to [UserName](Given the user name it path to correct team).''}

% Following are a few selected responses from developers discussing the activities which they want to see automated using AI. 

\begin{table*}
\newlength{\myboxheight}
\settoheight{\myboxheight}{1234567890\%}

\def\mybarchartx#1{
\resizebox {#1} {\myboxheight} {%
\begin{tikzpicture}[]
\definecolor{clr1}{RGB}{99,99,99}
\definecolor{clr2}{RGB}{240,240,240}
\begin{axis}[
      axis background/.style={fill=gray!10, draw=gray!50},
      axis line style={draw=none},
      tick style={draw=none},
      ytick=\empty,
      xtick=\empty,
      ymin=0, ymax=1, % this is 0.70 here, the other table is 0.60
      xmin=0, xmax=1]
\addplot [
      ybar interval=.5,
      fill=black,
      draw=none,
]
	coordinates {(1,1) (1,1)}; 
\addplot [
      ybar interval=.5,
      fill=black,
      draw=none,
]
	coordinates {(1,1) (0,1)}; 
\end{axis}%
\end{tikzpicture}%
}%
}

\centering
\newcommand{\mycategory}[1]{\emph{#1}}
\newcommand{\mybarchart}[2]{\mybarchartx{#1}~#2}
\newcommand{\mymidrule}[0]{\cmidrule{1-2}}
\caption{Which Tasks do Developers want to Automate?}
\begin{tabular}{@{}p{0.245\textwidth} p{0.6\textwidth} p{0.1\textwidth}@{}}  % Set fixed width for both columns
    \toprule
    \textbf{Category} & {\centering \textbf{Category description}} & \textbf{Count} \\
    \midrule

    \mycategory{Documentation} & Creating \& updating documentation, generating API docs from code comments, \& maintaining team knowledge bases. Efficiently understanding existing code, APIs, \& systems from documentation \& reports. & \mybarchart{41pt}{82} \\
    \mymidrule
    
    \mycategory{Environment set-up/maintenance} & Setting up SSH keys, installing software dependencies, syncing Git repositories, updating local branches, initializing new development instances/containers. & \mybarchart{33pt}{66}  \\
    \mymidrule

    \mycategory{Write/maintain tests} & Unit test creation for routine/similar coding tasks. Authoring, execution, \& monitoring of tests. & \mybarchart{30pt}{60} \\
    \mymidrule

    \mycategory{Task Tracking \& Backlog Management} & Task creation \& management in systems such as Azure DevOps, sub-tasks, repeatable tasks, or project status tracking. & \mybarchart{23.5pt}{47}\\
    \mymidrule

    \mycategory{Security \& Compliance} & Reviews, tracking, addressing issues (S360 tasks), handling security wave items, \& managing renewals or permissions. & \mybarchart{20pt}{40}\\
    \mymidrule

    \mycategory{Incident/Customer Issue Management} & Live site reporting, root cause analysis, incident correlation, triage, customer communication, and ICM handling. & \mybarchart{19pt}{38}\\
    \mymidrule 

    \mycategory{Communication} & Summarizing long email threads, creating ADO tasks, answering frequent technical questions from different developers. Efficient ways to handle meeting minutes, schedules, notes, action items, etc. & \mybarchart{16.5pt}{33}\\
    \mymidrule
    
    \mycategory{Deployment \& Release Management} & Deployments along with build-and-release pipelines often involve many manual steps and repetitive tasks. This includes validation checks, release notes creation, and understanding deployment health metrics. & \mybarchart{13pt}{26}\\
    \mymidrule    
    
    \mycategory{PR/Code Review/Change Management} & Code review feedback \& formatting, detection of anti-patterns, merge conflict resolutions, branch management, rerunning flaky jobs, merging development branches, handling pull request iterations, reviews, \& descriptions. & \mybarchart{12.5pt}{25}\\
    \mymidrule
    
    \mycategory{Infrastructure Monitoring \& Alerts} & The setup, maintenance, and interpretation of monitoring dashboards, alert rules, and other telemetry data. & \mybarchart{11.5pt}{23}\\
    \mymidrule
    
    \mycategory{Debugging}  & Automation around debugging processes, particularly surrounding build failures, errors, and runtime exceptions. & \mybarchart{7.5pt}{15}\\
    \mymidrule
    
    \mycategory{Code Refactoring} & Automating parts of the refactoring process, enforcing code quality standards, and detecting anti-patterns. & \mybarchart{6.5pt}{13}\\
    \mymidrule
    
    \mycategory{Build Automation} & Faster and reliable build systems to swiftly integrate and manage new code changes into existing codebases. & \mybarchart{5.5pt}{11}\\
    \mymidrule

    \mycategory{Maintenance of Internal Tools} & Maintenance tasks related to the upkeep of existing tools and systems. & \mybarchart{5pt}{10}\\
    \mymidrule

    \mycategory{Onboarding} & Adding new team members, giving them walk-throughs of the best practices, resources, setting up their development environment, and access permissions are areas suggested for improvement. & \mybarchart{3.5pt}{7}\\
    \mymidrule

    \mycategory{Cloud Infrastructure Maintenance} & Resource provisioning, scaling, performance monitoring, security patching, backups, cost optimization, network management, and incident response for efficient cloud operations. & \mybarchart{2.5pt}{5}\\
    \mymidrule
    
    \mycategory{Workflow Integration} & Better integration between version control systems, issue trackers, \& other development tools. & \mybarchart{1.5pt}{3}\\
    \mymidrule
    
    \mycategory{Reworking Legacy Code} & Assistance in dealing with codebase legacies, updating old systems, or adapting to new systems. & \mybarchart{1.5pt}{3}\\
    \bottomrule
\end{tabular}
\label{tab:AutomationTasks}
\end{table*}

%Insights box
\begin{tcolorbox}[
    colback=gray!10,  
    rounded corners,  
    boxrule=0.35mm,  
    left=5pt, right=5pt, 
    top=5pt, bottom=5pt 
]
\faLightbulbO\ \textbf{Takeaway}: 
Incorporating AI tools is crucial for enhancing developers' productivity, while also boosting their overall professional satisfaction. However, to maximize the benefits of automation, it is essential to align AI development efforts with the specific tasks and activities that developers most desire to automate. By focusing on these areas, research and automation initiatives can more effectively streamline workflows and address developers' real-world needs.

\end{tcolorbox}

% \subsection{Which tasks give the most satisfaction/dissatisfaction to the developers and their corresponding relation to cognitive-challenge}
% Doubt: Do we want to include this information, and try to link this to their productivity? Can remove

\section{Discussion}
\label{sec:discussion}

Our study provides valuable insight into the complex dynamics of a software developers' workweek. We aim to understand how the deviations from ideal and actual workweeks impact productivity and satisfaction levels. Additionally, we explored the role of up and coming AI tools and how we can modify our approach towards building new tools to build more tools that developers want. In this sections we will discuss a short summary of our findings and discuss the implications of our findings and impact on developers, managers and organizations in the software development industry as a whole.

\subsection{Time allotment vs productivity/satisfaction}

Our results show most developers generally have very different workweeks from their ideal ones. This deviation has a measurable impact on both productivity and satisfaction levels. We quantitatively showed that as the gap between the actual and ideal workweek increases there is a substantial negative impact on both productivity and satisfaction levels. This finding highlights the importance of having a better balance between what a developer wants to do and what they typically do in practice.

Unsurprisingly, activities that positively influence both productivity and satisfaction are those typically considered core development tasks, such as coding, documentation, and code refactoring. Conversely, time spent on maintaining and creating development environments, dealing with security and compliance issues, and excessive communication negatively impacts both metrics. These activity-level insights offer broader suggestions for how organizations can optimize their workflows to maximize time on core development activities while minimizing time on less preferred tasks.

These findings present several opportunities for software developers, teams, organizations and researchers working in this field. Organizations can potentially redesign or optimize existing workflows to better align with the needs of developers. There may also be opportunities to adjust team structures to distribute tasks more effectively, allowing developers to focus more on their preferred activities. Additionally, introducing new tools or improving existing ones could help streamline less desirable tasks, freeing up more time for core development activities.

A potential interesting direction for future research would be to explore the long-term impact of better aligning developers' workweeks with their ideal ones. This could involve examining changes in code quality, project success metrics, and overall user experience over time. However, it's important to note that finding the right balance will likely require ongoing assessment and adjustment, as the ideal workweek may vary not only between individuals but also evolve over time for each developer.

% Perhaps, unsurprisingly activities that positively influence both productivity and satisfaction are things that are typically considered as core development activities like coding, documentation, code refactoring. On the other end, time spent on maintaining and creating development environments, dealing with security and compliance issues and excessive communication negatively impacts both the metrics. These activity level insights offer wider suggestions how companies can optimize their workflows to maximize time on core development activities and perhaps less time on the other activities. There may be opportunities to redesign workflows, adjust team structures, or implement new tools to better align with individual developers' ideal workweeks.

% Quantify using SPACE/Truce framework, how different people will have different workweeks. Balancing the needs 

% How other activities affect the long-term affects of softtware development

\subsection{Future impact of AI tools}

We observed a strong relation between how frequently developers used AI tools and their reported satisfaction and productivity levels. Developers who use these tools daily tended to report the highest level of both satisfaction and productivity. This underscores the importance of these tools in the modern software development landscape.

As discussed earlier, developers most want to see activities automated, which we observed to correlate with lower satisfaction and higher productivity levels. These findings present a new exciting direction for future AI tools and automation development efforts. By focusing the development of AI tools on areas which can have the dual benefit of boosting developer productivity and satisfaction whilst at the same time allowing them to focus on tasks they enjoy and find more fulfilling, organizations can foster a more motivated and effective workforce.

Currently a large number of AI tools focus specifically on helping developers write code ~\cite{githubcopilot, cursor, cody}. However, our findings suggest a strong desire for automation in non-coding related tasks such as documentation, task management, security and compliance, and communication. This reveals a gap in the current AI tool landscape and presents a significant opportunity for innovation. Developing new AI tools that address these areas could offer substantial benefits not only to individual developers but also to organizations as a whole.

The potential impact of such targeted AI tool development is far-reaching. By automating tasks that developers find less productive or satisfying, we could see a shift in how developers allocate their time, potentially bringing their actual workweeks more in line with their ideal ones. This could lead to higher job satisfaction, increased productivity, and potentially even improvements in code quality and project outcomes as developers are able to focus more on core development tasks they find most engaging. However, as the adoption of these tools grow, there might be a change in what a developer considers to be an ideal workweek. Future research should not only focus on developing these tools but also on understanding their broader impacts on the software development as a whole.

As we identify automation candidates that developers perceive as associated with lower satisfaction but high productivity levels, it is crucial to consider the diversity of roles and workloads across the team. For instance, team leads (TLs) often have workloads that differ significantly from those of individual contributors, due to the inherent responsibilities of their roles. A potential impact on satisfaction could stem from these deviations, where some developers may feel disconnected from their regular or prior tasks--tasks they found fulfilling or technically engaging. At the same time, there may be developers who view this shift as an opportunity to explore new activities, broaden their skill sets, or maintain cognitive engagement in their workweek.
This duality in perspectives present an interesting avenue for future researchers. Future studies could investigate how these workload deviations influence both productivity and satisfaction--particular in a role-specific context--and how these insights can inform the selection of tasks for automation.

\section{Related Work}
\label{sec:related_work}

% \sukrit{Very very rough related work pass, PTAL. I'll add more things 10/9 night EST time, slightly busy with a few other things.}

To our knowledge, this is the first work that compares a developer's ideal and typical workweek and assesses the impact of their deviations on productivity and satisfaction. We also explore the role of AI tools in the software development process and capture potential future directions for AI tool development. We give a brief summary of existing work in three areas: developer activities and time spent, developer productivity and satisfaction, and AI tools and automation. 

\subsection{Developer activities and time allotment}
Understanding how developers spend their time across various activities has been a very well explored topic. Prior work analyzed how developers spend their time within the IDE by specifically tracking IDE usage, monitoring test and other refactoring feature usage \cite{i_know_what_you_did,vs_code_tracking}. This work did not focus on other activities outside the IDE like collaboration, communication and planning.

More recent works have captured a more holistic picture of what a developer does and aimed to capture the time spent through various methods of capturing data like surveys \cite{8666786,beller-software,activitiesTime2}, observations and interviews~\cite{gonccalves2011collaboration}, and through tracking usage of the computer~\cite{vs_code_tracking, i_know_what_you_did}. While many studies focus on understanding the daily activities, fewer have looked at patterns across an entire workweek. Our study builds on this gap, extending our window to the past work week allowing us to get a more comprehensive understanding of how a developer spends their time and also reducing the impact of day to day variations. In addition, we compare the actual workweek with an ideal workweek.

% There have been works like \cite{8666786} which provided a comprehensive categorization of different activities, instead of relying exclusively on this we also 
% Our work builds further upon this, not only analyzing the time spent on different activities over a week but also by these activities are related to self-reported productivity and satisfaction levels as well. We additionally, also look at the deviations between the actual and expected workweek of a developer.

\subsection{Developer productivity and satisfaction}
There exists a substantial amount of prior work on trying to understand the factors that influence developer productivity and satisfaction~\cite{storey2019towards,space,murphyhill-tse,sadowski-book}. There is significant work in understanding how different factors like interruptions~\cite{mark2008cost, bailey2001effects}, emails and communication ~\cite{houck2023the} impact both productivity and satisfaction.

Prior work~\cite{graziotin2014happy, graziotin2014software, kim2019understanding, emotion_perceive_productivity} showed also a link between developer happiness and increased productivity. Therefore, it is important to analyze how different activities affect a developers satisfaction and in-turn productivity. Masood et al.~\cite{like_dislike} analyzed how different tasks and activities affect developer happiness and productivity. 

% \sukrit{Rewrite paragraph. Also need to include mind the gap paper}
% There are a few lines of work \cite{graziotin2017unhappiness} that indicate the correlation between a developer having a good workday and the tasks in the day typically aligning with the expected with the role. 

\subsection{AI tools and automation in software development}

The rapid development and adoption of AI tools and automation in software development has opened new avenues for understanding their impacts on the field. There have been works~\cite{copilot_prod, borg2024does, ulfsnes2024transforming, peng2023impact} that analyze how different AI tools impact developer productivity. There is also a body of work~\cite{khemka2024towards,bird2022taking} on understanding specific tasks that developers want to use AI tools. 

We build upon these existing studies by analyzing the impact of AI tool usage on productivity and satisfaction. We further investigate whether developers want particular tasks to be automated. This can drive future development of AI tools and ensure that they are aligned with the actual requirements of the developers. 

\section{Conclusion}
\label{sec:conclusion}

In this paper, we provide valuable insights on the complex relationship between software developers' workweek and how it's deviation from their ideal workweek affects their productivity and satisfaction levels. Here are our key findings: 

\begin{enumerate}
    \item There is a large deviation between the ideal and actual workweeks, with developers clearly wanting to spend more time on core developmental activities and less on communication/maintenance ones.
    \item As the deviation between the ideal and actual workweek increases the productivity and satisfaction levels tend to fall.
    \item Usage of AI tools by a software developer is positively correlated with higher levels of productivity and satisfaction. This effect is especially more pronounced when the frequency of usage increases.
    \item Developers have strong preferences towards automation on activities like documentation, environment setup, testing, monitoring etc. These insights will inform the research and development of future AI tools.
\end{enumerate}

The insights we present in this paper have important implications for developers, managers and companies as a whole. Teams and companies can potentially boost satisfaction and productivity levels by working on more closely aligning a typical workweek with a developer's ideal workweek. Furthermore, the development of future AI tools can take into account the needs of a developer as discussed in our paper and fill gaps in this area. 

While our work is restricted to Microsoft, it would be interesting to see how our research fits into different organizational contexts and over longer intervals of time. Additionally, another interesting aspect to look at would be how AI-tools potentially change a typical and ideal workweek of a developer as the adoption and scope of AI increases.

In conclusion, we try to understand how time spent on different activities over a workweek and the deviation from a developers ``ideal" workweek impact productivity and satisfaction levels. 
Our findings reveal significant potential to reshape, innovate, and streamline workflows, enhancing developer productivity and satisfaction. This will lead to more fulfilling work experiences and higher quality outcomes, benefiting both developers and the organizations.

\section*{Acknowledgments}
We thank our survey participants for their participation. We also thank the exploratory interview participants for helping guide our study design.
We thank the ICSE reviewers for their insightful and constructive comments.

% \pagebreak
% \balance
\bibliographystyle{IEEEtran.bst}
\bibliography{refs}

\end{document}